\documentclass[12pt]{article}
\usepackage{epsfig}
\makeatletter
\def\alt{\mathrel{\mathpalette\gl@align<}}
\def\agt{\mathrel{\mathpalette\gl@align>}}
\def\gl@align#1#2{\lower.6ex\vbox{\baselineskip\z@skip\lineskip\z@
\ialign{$\m@th#1\hfil##\hfil$\crcr#2\crcr\sim\crcr}}}
\makeatother

\begin{document}
\begin{flushright}
{\tt hep-ph/0301014}\\
OSU-HEP-03-01 \\
January, 2003 \\
\end{flushright}
\vspace*{2cm}
\begin{center}
{\baselineskip 25pt
\large{\bf 
Gauge Higgs Unification in the Left-Right Model
}}

\vspace{1cm}

{\large 
Ilia Gogoladze\footnote
{On a leave of absence from: Andronikashvili Institute of Physics, GAS, 380077, Tbilish, Georgia.\\
email: {\tt ilia@hep.phy.okstate.edu}}, 
Yukihiro Mimura\footnote
{email: {\tt mimura@hep.phy.okstate.edu}}
and 
S. Nandi\footnote
{email: {\tt shaown@okstate.edu}}
}
\vspace{.5cm}

{\small {\it Physics Department, Oklahoma State University, \\
             Stillwater, OK 74078}}

\vspace{.5cm}

\vspace{1.5cm}
{\bf Abstract}
\end{center}

We construct a supersymmetric left-right model in four dimension with gauge-Higgs unification
starting from a $SU(3)_c \times SU(4)_w \times U(1)_{B-L}$ gauge symmetry in five dimension.
The model has several interesting features, such as, 
the CKM mixings in the quark sector are naturally small while for the neutrino
sector it is not,
light neutrino masses can be generated via the seesaw mechanism in the usual way,
and the model has a $U(1)_R$ symmetry which naturally forbid dimension five proton decay operators.
We also discuss the grand unification of our model
in $SO(12)$ in five dimensions.

\thispagestyle{empty}

\bigskip
\newpage

\addtocounter{page}{-1}

\section{Introduction}
\baselineskip 20pt

Recent topics of the theories in higher dimensions give us a lot of interesting
phenomenological pictures.
One of the most attractive motivations of extension of dimensions
is that the variety of particles in Nature
can be understood by means of a geometrical language.
For example, gauge fields with the coordinate for the extra dimensions
behave as scalar fields in 4 dimension.
Since masses of the gauge bosons are prohibited by gauge invariance,
the scalar field originated from gauge bosons can be a good candidate 
of the low energy Higgs fields, which breaks electroweak symmetry.
That leads to the idea of the gauge-Higgs unification in the 
higher dimensional theories \cite{Manton:1979kb,Hosotani:1983xw,Hall:2001tn}.
Recent realization of the phenomenological models in higher dimensions
makes us encourage to revisit the idea \cite{Krasnikov:dt,Hall:2001zb,Burdman:2002se,Haba:2002vc}.

We consider that the extra dimensions are compactified 
in an orbifold
in order to make chiral theories in 4D,
since 5D fermions include both chirality in 4D language.
In an orbifold space, such as $S^1/Z_2$,
we can impose boundary conditions at the folding places,
and the gauge
symmetry can be broken through these boundary conditions \cite{Hosotani:1983xw,Scherk:1978ta}.
Recently, a great deal of works has been done on
the gauge symmetry breaking using the orbifold boundary conditions,
and these lead to many attractive features of the unified gauge theories 
in higher dimensions \cite{Kawamura:1999nj,Hebecker:2001jb}.
Using the orbifold boundary condition, we can project out unwanted fields
such as colored Higgs triplets in the grand unified theories \cite{Kawamura:1999nj}.
In such a progress of the higher dimensional unified theories,
interesting ideas of gauge-Higgs unification are suggested.
In the higher dimensional supersymmetric theories,
the gauge multiplet contains both vector multiplet and 
chiral supermultiplet in 4D.
Assigning the different $Z_2$ parity between vector multiplet 
and chiral supermultiplets,
we can make vector multiplet massless but chiral supermultiplets heavy,
which means the supersymmetry is broken.
If we break gauge symmetry through boundary condition simultaneously,
a part of the chiral supermultiplets can have a zero mode
which remains massless in the low energy.
Then, we can identify such a supermultiplet with the low energy Higgs field.
This is the main idea of the gauge-Higgs unification which we consider in this paper.
This idea was realized in the 6D $N=2$ supersymmetric theories \cite{Hall:2001zb},
and more recently in the 5D $N=1$ supersymmetric theories \cite{Burdman:2002se,Haba:2002vc}.
The latter scenario gives us an interesting possibility
that the gauge and Yukawa coupling constants have the same origin.
Since the Yukawa interactions arise from the gauge interaction
in the 5D lagrangian,
those two coupling constants are ``unified" in the 5D theory.
This is a very interesting feature of the 5D gauge-Higgs unified scenario.
In the Ref.\cite{Burdman:2002se}, the authors considered the theory of $SU(3)_w$ and $SU(6)$
as an example of the scenario,
but in their models, there are many unwanted fields in the matter hypermultiplets.
In order to make the unwanted fields heavy, they need many brane-localized fields
for each generation.
The brane fields are actually needed to cancel the gauge anomaly in the 4D theory
which arises from the zero modes of bulk hypermultiplets,
and that means it is not easy to understand the anomaly free structure in their models.
In Ref.\cite{Haba:2002vc}, the authors consider the gauge-Higgs unification in larger gauge group
such as $E_6$, $E_7$ and $E_8$.

In this paper,
we construct a supersymmetric left-right model,
$SU(3)_c \times SU(2)_L \times SU(2)_R \times U(1)_{B-L}$ in 4D \cite{Pati:1974yy,Mohapatra:1974hk}
with gauge-Higgs unification.
We point out that we are unifying the gauge fields with only the Higgs multiplets which give masses
to the standard model fermions.
We start with a supersymmetric model
with the gauge group $SU(3)_c \times SU(4)_w \times U(1)_{B-L}$
in 5D.
The $SU(4)_w$ gauge symmetry is broken down to $SU(2)_L \times SU(2)_R \times U(1)_X$ 
by orbifold boundary condition,
and then the gauge symmetry in 4D become left-right symmetric gauge group
with extra $U(1)_X$ symmetry
($U(1)_X$ is broken to nothing using suitable brane interactions).
In this model, no unwanted zero modes arise from the matter hypermultiplets.
It is easy to see the anomaly free structure, and this structure 
naturally give rise to even number of families
of the matter hypermultiplets.
The left-right symmetric construction gives us a good picure to the 
scenario of the 5D gauge-Higgs unification.
Our model naturally leads to small CKM mixings in the quark sector, 
while in the neutrino sector, mixing can be large.
The model is nicely grand unified in $SO(12)$ in 5D.

Our paper is organized as follows:
In section 2, we construct our supersymmetric $SU(4)_w$ model in 5D
with gauge-Higgs unification and show how orbifold compactification
leads to left-right symmetric model in 4D. 
In section 3, we discuss the quark and lepton mass matrices
and mixings in our model and various other features.
Grand unification of our model in $SO(12)$ is contained in section 4.
Section 5 has our conclusions and discussions.


\section{Gauge Higgs Unification in $SU(4)_w$ Gauge Theory}

In this section, we discuss the construction of the gauge-Higgs unification 
in 5D $N=1$ supersymmetric theory based on the gauge group $SU(3)_c \times SU(4)_w \times U(1)_{B-L}$.
We will consider $S^1/Z_2$ orbifold,
which is constructed by identifying the coordinate of the fifth dimension, $y$,
under two parity transformations:
$Z_2: y \rightarrow -y$ and $Z_2^\prime : y^\prime \rightarrow -y^\prime$, 
where $y^\prime = y + \pi R$.
The orbifold space is regarded as a interval $[0,\pi R]$
and 4 dimensional walls (we call 4D wall brane) are placed at the folding point $y=0$ and $y=\pi R$.
The 5D $N=1$ supersymmetric theory corresponds to 4D $N=2$ supersymmetric theory.
In 4D language,
the $N=2$ gauge multiplet contains one $N=1$ vector multiplet $V (A_\mu,\lambda)$
and one $N=1$ chiral multiplet $\Sigma (\sigma+iA_5,\lambda^\prime)$.
The boundary conditions at 4D walls are given as
\begin{equation}
\left( \begin{array}{c}
         V \\ \Sigma
       \end{array}
\right) (x^\mu, -y) =
\left( \begin{array}{c}
         P V P^{-1} \\ -P \Sigma P^{-1}
       \end{array}
\right) (x^\mu, y),
\quad
\left( \begin{array}{c}
         V \\ \Sigma
       \end{array}
\right) (x^\mu, -y^\prime) =
\left( \begin{array}{c}
         P^\prime V P^{\prime-1} \\ -P^\prime \Sigma P^{\prime-1}
       \end{array}
\right) (x^\mu, y^\prime),
\label{boundary-condition}
\end{equation}
where $P$ and $P^\prime$ acts on gauge space.
Then 4D $N=2$ supersymmetry is broken down to $N=1$ supersymmetry
and non-trivial $P$ and $P^\prime$ breaks gauge symmetry $G$ down to $H$.
In the case that $P=P^\prime$,
the vector multiplets has $Z_2 \times Z_2^\prime$ parity
as $(+,+)$ for unbroken gauge symmetry, and $(-,-)$ for broken one $G/H$. 
On the other hand, for the chiral multiplet, the signature of the parity is opposite.
Since only $(+,+)$ components have massless modes,
the chiral multiplet $\Sigma$ for the broken generator remains massless.
We identify the massless chiral multiplet as Higgs field to break electro-weak symmetry
$SU(2)_L \times U(1)_Y$.

We will apply this 5D gauge-Higgs unification scenario to left-right model.
We consider $SU(3)_c \times SU(4)_w \times U(1)_{B-L}$ symmetry
as the bulk gauge symmetry.
We need this $U(1)_{B-L}$. The reason is that after the orbifold breaking of $SU(4)_w$,
although we get an $U(1)_X$, it does not have the right $B-L$ quantum numbers as required
by the bi-doublet Higgs in the left-right theory.
The boundary condition breaks $SU(4)_w$ down to $SU(2)_L \times SU(2)_R \times U(1)_X$
on the 4D walls,
if we use $P=P^\prime = {\rm diag} (1,1,-1,-1)$.
The $SU(4)$ adjoint is decomposed as
\begin{equation}
\mathbf{15} = \mathbf{(3,1)}_0 + \mathbf{(1,3)}_0 + \mathbf{(1,1)}_0 + \mathbf{(2,2)}_2 + \mathbf{(2,2)}_{-2},
\end{equation}
where the numbers in the subscripts represent $U(1)_X$ charges.
The two bi-doublets correspond to the unbroken generator,
and their $\Sigma$ components of the gauge multiplet
remain massless after the compactification. These can be identified to be the bi-doublet Higgs
required to gives masses to the fermions in the left-right symmetric theory, and in our model,
they originate from the gauge supermultiplet.
We define the hypercharge as $Y \equiv T_R^3 + (B-L)/2$,
where $T_R^3$ is a generator for $SU(2)_R$.
The $B-L$ charge of the bi-doublets are zero
since those ones come from $SU(4)_w$ gauge multiplet.
Thus those bi-doublet quantum numbers are same as the ones needed in the left-right model.
Since we don't need the $U(1)_X$ symmetry at low energy,
we will break $U(1)_X$ without mixing with $SU(2)_R$ and $U(1)_{B-L}$.
To do that, we add the brane fields $z$ and $\bar z$, which have $+1$ and $-1$ $U(1)_X$ charge
and are singlets for the other symmetry.

Next we consider the matter fields.
The matter fields should be bulk fields in the scenario of 5D gauge-Higgs unification
since the chiral superfield $\Sigma$ transforms non-linearly under 5D gauge transformation
and we cannot make Yukawa coupling with the chiral field $\Sigma$ at 4D walls.
So the matter fields are $N=2$ hypermultiplets $(\Psi, \Psi^c)$,
where the $\Psi$ and $\Psi^c$ are the $N=1$ chiral multiplets.
If we take this non-conjugated field $\Psi$ as fundamental representation under $SU(4)_w$
(conjugated field $\Psi^c$ is the anti-fundamental),
then $\Psi$ and $\Psi^c$ are decomposed as
\begin{eqnarray}
\Psi &=& \mathbf{(2,1)}_1 + \mathbf{(1,2)}_{-1}, \\
\Psi^c &=& \mathbf{(2,1)}_{-1} + \mathbf{(1,2)}_{1}. 
\end{eqnarray}
The boundary condition is givn as follows:
\begin{equation}
\Psi (x^\mu, -y) = s P \Psi (x^\mu, y), \quad \Psi^c (x^\mu, -y) = - s P \Psi^c (x^\mu, y),
\end{equation}
\begin{equation}
\Psi (x^\mu, -y^\prime) = s P^\prime \Psi (x^\mu, y^\prime), \quad \Psi^c (x^\mu, -y^\prime) = - s P^\prime \Psi^c (x^\mu, y^\prime),
\end{equation}
where $s=\pm1$.
In the case where $s=1$, $\mathbf{(2,1)}_1$ and $\mathbf{(1,2)}_1$ have $(+,+)$ parity and others have $(-,-)$,
and in the case where $s=-1$, the parity charges are opposite.
With this boundary condition, we obtain the left- and right-handed quarks and leptons assigning appropriate
color and $B-L$ charges.
Choosing $s=1$, we have one family of quarks and leptons, which are the massless,
in the representations $(SU(3)_c,SU(2)_L,SU(2)_R)_{U(1)_{B-L},U(1)_X}$:
\begin{equation}
Q_L : \mathbf{(3,2,1)}_{1/3,1}, \quad Q_R^c : \mathbf{(3^*,1,2)}_{-1/3,1}, \quad
\ell_L : \mathbf{(1,2,1)}_{-1,1}, \quad \ell_R^c : \mathbf{(1,1,2)}_{1,1}.
\end{equation}
It is important to notice that gauge anomaly with respect to $U(1)_X$ arises
such as $SU(3)_c^2 \times U(1)_X$, $SU(2)_{L,R}^2 \times U(1)_X$, and $U(1)_X^3$.
To cancel the anomaly, we need one more family with $s=-1$.
In other words, fixing $s=1$, we need both $(\mathbf{3,4})_{1/3}$ and $(\mathbf{3,4^*})_{1/3}$ 
for quark non-conjugated fields, 
and need both $(\mathbf{1,4})_{-1}$ and $(\mathbf{1,4^*})_{-1}$ for lepton non-conjugated fields
in the representation $(SU(3)_c, SU(4)_w)_{U(1)_{B-L}}$.
This gives an interesting feature of this model:
The number of the generation is even.
We note that it is possible that the $U(1)_X$ anomaly can be canceled by introducing appropriate brane fields.
However, since we have to make those brane fields heavy by folding Dirac mass with another multiplets,
this possibility is not economical.
This $U(1)_X$ gauge anomaly gives us a strong constraint to construct a model.

We now discuss the coupling among the matter fields $\Psi$, $\Psi^c$ and the chiral multiplet $\Sigma$.
The action is written as \cite{Arkani-Hamed:2001tb}
\begin{equation}
S_{5D}= \int d^4 x dy \left[ \int d^4 \theta (\bar \Psi e^{-V} \Psi + \Psi^c e^V \bar \Psi^c)
+ \left(\int d^2 \theta \Psi^c (\partial_5 - \frac1{\sqrt2} \Sigma ) \Psi + {\rm h.c.} \right) \right],  
\end{equation}
and the ``superpotential" term gives us the Yukawa coupling in the 4D theory.
Naming the $\mathbf{(2,2)}_{-2}$ and $\mathbf{(2,2)}_{2}$ in the $\Sigma$ multiplet as
$\Phi_1$ and $\Phi_2$ respectively,
we obtain the following Yukawa coupling for quark:
\begin{equation}
S_{4D} = \int d^4 x  \int d^2 \theta (y_1 Q_{+L}^{(0)} \Phi_1^{(0)} Q_{+R}^{c(0)} + y_2 Q_{-L}^{(0)} \Phi_2^{(0)} Q_{-R}^{c(0)}) + {\rm h.c.},
\label{yukawa}
\end{equation}
where we denote that $Q_{\pm L}$ and $Q_{\pm R}^c$ having $\pm 1$ charge for $U(1)_X$ symmetry,
and the subscript $(0)$ as the zero modes. 
Yukawa couplings for leptons are also written in the same way. 
We can see that the Yukawa coupling constants $y_1$ and $y_2$ originate from gauge coupling,
and thus, the Yukawa and gauge couplings have the same origin.
This is the most interesting feature of the 5D gauge-Higgs unification scenario.
However, then we find that the Yukawa couplings for all the families have
the same values and equal to the gauge coupling.
Observation excludes such a situation for the first and second family, and we need to solve the problem.
The problem can be solved by introducing bulk masses for the hypermultiplets $\Psi$ and $\Psi^c$ such as
\begin{equation}
S = \int d^4 x dy \left[\int d^2 \theta M (y) \Psi^c \Psi + {\rm h.c.} \right],
\end{equation}
where $M(y) = M ( \epsilon(y) + \epsilon (\pi R - y) )/2$ and $\epsilon(y)$ is a step function.
With this bulk mass, the zero-mode wave functions of the fermions localize at 4D walls.
If we choose $M<0$, the zero-mode wave function of $\Psi$ localizes at $y=0$ with a profile $e^{-|M|y}$,
and for $\Psi^c$, the zero-mode localizes at $y=\pi R$ with profile $e^{|M|(y-\pi R)}$.
Then we find that the Yukawa couplings are different from 4D gauge coupling $g$ such as
\begin{equation}
y = \frac{\pi R |M|}{\sinh (\pi R |M|)}\ g.
\end{equation}
Therefore, if $|M|R$ is larger than 1, the Yukawa couplings are exponentially suppressed,
and thus we can obtain the hierarchical fermion masses.  

In this section, we have constructed the 5D gauge-Higgs unification model
in the context of our $SU(4)_w$ gauge theory.
In this 5D scenario, not only the low energy Higgs fields come from gauge multiplet,
but also the Yukawa couplings originate from gauge interaction.
The fermion mass hierarchy is generated by the bulk masses of the matter fields.
But still, we don't have generation mixing for the quarks since the Yukawa couplings 
come from gauge interactions which do not produce such mixings.
We will see how the family mixings are generated 
in the next section.

\section{Quark and Lepton Mass Matrices}

In this section, we will investigate the quark and lepton mass matrices of our $SU(4)_w$ model.
We recall that number of generation of this model should be even number
because of $U(1)_X$ gauge anomaly.
We will consider 4-generation model here.
The two of the 4 generations have $+1$ charge for $U(1)_X$ and the others have $-1$ charge.

We introduce chiral superfields $\bar Q_L$, $\bar Q_R^c$, $\bar \ell_L$ and $\bar \ell_R^c$ on the 4D walls.
The quantum numbers $(SU(3)_c,SU(2)_L,SU(2)_R)_{U(1)_{B-L},U(1)_X}$ of those fields are the following:
\begin{equation}
\bar Q_L(\mathbf{3^*,2,1})_{-1/3,0},\quad \bar Q_R^c(\mathbf{3,1,2})_{1/3,0},
\quad \bar \ell_L(\mathbf{1,2,1})_{1,0}, \quad \bar \ell_R^c(\mathbf{1,1,2})_{-1,0}.
\end{equation}
Then we will have the following superpotential terms for quark:
\begin{eqnarray}
S \!\!&=&\!\! \int d^4x dy (\delta(y)+\delta(y-\pi R)) \times  
\label{brane-int} \\ 
&& \!\! \left[\int d^2 \theta \left( (z Q^1_L + \bar z Q^2_L + z Q^3_L + \bar z Q^4_L) \bar Q_L   
+ (z Q^{c1}_R + \bar z Q^{c2}_R + z Q^{c3}_R + \bar z Q^{c4}_R) \bar Q_R^c \right)
+ {\rm h.c.}\right], \nonumber 
\end{eqnarray}
where $z$ and $\bar z$ are the chiral superfields which have $\pm1$ $U(1)_X$ charge and singlet
for the other symmetry.
The superscripts 1,2,3,4 denote the family indices.
We also have similar superpotential terms for leptons.
The vacuum expectation values for $z$ and $\bar z$ break $U(1)_X$ symmetry,
generate flavor mixings and make the 4th family heavy.
{}From Eqs.(\ref{yukawa}) and (\ref{brane-int}), the $5\times5$ quark mass matrices 
for left-handed quarks $(Q_L^1,Q^2_L,Q^3_L,Q^4_L,\bar Q^c_R)$ and 
right-handed quarks $(Q_R^{c1},Q^{c2}_R,Q^{c3}_R,Q^{c4}_R,\bar Q_L)$
are obtained to be
\begin{equation}
{\cal M}_{u,d} = \left( \begin{array}{cc}
                         M_{u,d} & \begin{array}{c} a_1 \\ a_2 \\ a_3 \\ a_4 \end{array} \\
                         \begin{array}{cccc} a_1^\prime & a_2^\prime & a_3^\prime & a_4^\prime 
                         \end{array} &0
                        \end{array} \right), \quad
M_{u,d} = {\rm diag} (y_1 \kappa_1^{u,d}, y_2 \kappa_2^{u,d}, y_3 \kappa_1^{u,d}, y_4 \kappa_2^{u,d}),
\label{mass-matrix}
\end{equation}
where $\kappa_a^{u,d}$ $(a=1,2)$ are the vacuum expectation values of $\Phi_a$ such as
\begin{equation}
\langle \Phi_a \rangle = \left( \begin{array}{cc}
                                \kappa^u_a & 0 \\
                                0 & \kappa^d_a 
                                \end{array}
                         \right),
\end{equation}
and $a_i$ and $a_i^\prime$ are the vacuum expectation values of $z$ and $\bar z$ multiplied by 
order 1 factors.
There is an ambiguity how to select which generation gets mass by $\Phi_1$ or $\Phi_2$, 
but two of the generations should get masses by $\Phi_1$ and the other two by $\Phi_2$.
The parameters $a_i$ $(a_i^\prime)$ for the up-type and down-type quarks are same because 
up and down-type quarks are in the same multiplets $\bar Q_L$ $(\bar Q^c_R)$.
If we assume the left-right symmetry $Q^i_L \leftrightarrow Q^{ci*}_R$
and $\bar Q_L \leftrightarrow \bar Q_R^{c*}$ in the same way as ordinary 4D left-right symmetric model,
we have $a_i^\prime = a_i^*$ and the mass matrices are hermitian.

We now calculate the mass eigenvalues
of the mass matrices givne in Eq.(\ref{mass-matrix}).
We assume that $a_i$'s are much larger scale than $y_i \kappa_a^{u,d}$.
Then, one of the 4 generations decouples around $U(1)_X$ breaking scale
and the other 3 generations remain at weak scale.
We find that weak scale quark masses, assuming $y_1 \kappa_1$, $y_2 \kappa_2 \ll y_3 \kappa_1 \ll y_4 \kappa_2$,
\begin{equation}
m_{t,b} = \sqrt{\frac{|a_1|^2+ |a_2|^2 + |a_3|^2}{|a_1|^2+ |a_2|^2 + |a_3|^2 + |a_4|^2}}
          \sqrt{\frac{|a_1^\prime|^2+ |a_2^\prime|^2 + |a_3^\prime|^2}{|a_1^\prime|^2+ |a_2^\prime|^2 + |a_3^\prime|^2 + |a_4^\prime|^2}} \ y_4 \kappa_2^{u,d}, 
\end{equation}
\begin{equation}
m_{c,s} = \sqrt{\frac{|a_1|^2+ |a_2|^2}{|a_1|^2+ |a_2|^2 + |a_3|^2}}
          \sqrt{\frac{|a_1^\prime|^2+ |a_2^\prime|^2}{|a_1^\prime|^2+ |a_2^\prime|^2 + |a_3^\prime|^2}} \ y_3 \kappa_1^{u,d}, 
\end{equation}
\begin{equation}
m_{u,d} = \left| \frac{a_1^*}{\sqrt{|a_1|^2+ |a_2|^2}}
          \frac{a_1^\prime}{\sqrt{|a_1^\prime|^2+ |a_2^\prime|^2}} \ y_2 \kappa_2^{u,d}
        + \frac{a_2^*}{\sqrt{|a_1|^2+ |a_2|^2}}
          \frac{a_2^\prime}{\sqrt{|a_1^\prime|^2+ |a_2^\prime|^2}} \ y_1 \kappa_1^{u,d} \right|.
\end{equation}
Since the mass discrepancy of the up and down-type quarks comes from the difference
of $\kappa_a^u$ and $\kappa_a^d$,
the masses of third and second family should come from different $\Phi_a$;
otherwise, unacceptable relation appears such as $m_c/m_t = m_s/m_b$.
The similar relation between first and second generation (or first and third generation) can appear,
so we have assumed $y_1 \kappa_1 \sim y_2 \kappa_2$.

Next we investigate the Cabibbo-Kobayashi-Maskawa (CKM) angles.
We define the $4\times 4$ unitary matrix $U$ and $U^\prime$ such that
\begin{equation}
(a_1, a_2, a_3, a_4) U = (0,0,0,a), \quad 
(a_1^\prime, a_2^\prime, a_3^\prime, a_4^\prime) U^\prime = (0,0,0,a^\prime),
\end{equation}
where $a=\sqrt{\sum_i |a_i|^2}$, and $a^\prime = \sqrt{\sum_i |a_i^\prime|^2}$.
The unitary matrices mix the families largely.
However, since 
these mixing matrices are the same for the
up and down-type quarks,
the large mixings don't contribute to the CKM matrix.
After removing the non-physical large mixing,
we have the quark mass matrices:
\begin{equation}
{\cal M}^\prime_{u,d} = 
                        \left( \begin{array}{c|c}
                         U^\dagger M_{u,d} U^\prime & \begin{array}{c} 0 \\ 0 \\ 0 \\ a \end{array} \\ \hline
                         \begin{array}{cccc} 0& 0 & 0 & a^\prime 
                         \end{array} &0
                        \end{array} \right).
\end{equation}
Having 4-th generation decoupled, we obtain
the 3-generation quark mass matrices:
\begin{equation}
(\mathbf{m}_{u,d})_{ij} = [(U^\dagger M_{u,d} U^\prime)_{ij}]_{i,j=1,2,3}.
\end{equation}
We can easily calculate the expression of the matrices explicitly, but here we
show the approximate structure of the matrices:
\begin{equation}
\mathbf{m}_{u,d} \sim \left( \begin{array}{ccc}
                             m_{u,d} & m_{u,d} & m_{u,d} \\
                             m_{u,d} & m_{c,s} & m_{c,s} \\
                             m_{u,d} & m_{c,s} & m_{t,b}
                             \end{array} \right).
\label{appro-mass}
\end{equation}
{}From Eq.(\ref{appro-mass}), we obtain the CKM angles approximately:
\begin{equation}
V_{us} \simeq O(1) \ \frac{m_d}{m_s}, \quad V_{cb} \simeq O(1) \ \frac{m_s}{m_b}, \quad 
V_{ub} \simeq O(1) \ \frac{m_d}{m_b}.
\end{equation}
The $O(1)$ factors are the functions of the parameters $a_i$ and $a_i^\prime$.
We spoil the good relation for Cabibbo angle $V_{us} \simeq \sqrt{m_d/m_s}$,
but still we can fit all the CKM angles and quark masses to the observed values
by choosing the parameters
such as $a_i$ and $a_i^\prime$.
It is worth noting here that smallness of the CKM angle is automatically derived
if we have hierarchical Yukawa coupling in the diagonal elements.
We don't need to assume the hierarchical coupling in the brane interaction in Eq.(\ref{brane-int}),
thanks to unification of up and down-type quarks via the gauge-Higgs unification.
Furthermore,
since we use the exponential profile of the fermion localization to derive the
Yukawa hierarchy,
we do not need any small numbers in the fundamental theory to derive the
quark mass structure.

Now we briefly discuss the lepton sector.
We can construct the charged lepton mass matrix and Dirac neutrino mass 
in a way similar to the quark sector.
To have the Majorana mass for the right-handed neutrino,
we introduce $SU(2)_R$ triplet Higgses $\Delta$ $\mathbf{(1,1,2)}_{2,0}$ and 
$\bar \Delta$ $\mathbf{(1,1,2)}_{-2,0}$ whose vacuum expectation values
break $SU(2)_R \times U(1)_{B-L}$ down to $U(1)_Y$.
Giving appropriate coupling permitted by $U(1)_X$ symmetry,
we introduce the brane superpotential term at $y=0$,
\begin{equation}
\delta (y) \left[\int d^2 \theta (f_{ij} \bar \Delta \ell_R^{ci} \ell_R^{cj} + \Delta \bar \ell_R^{c} \bar \ell_R^{c})
 +{\rm h.c.}\right],
\end{equation}
where some elements of the coefficient $f_{ij}$ should be zero because of $U(1)_X$ symmetry.
The brane superpotential term generates Majorana mass term for the right-handed neutrino in 4D,
and gives small neutrino masses for the light neutrinos through the seesaw mechanism.
If the right-handed leptons are localized at $y=0$ by appropriate bulk masses,
the Majorana mass matrix does not have hierarchical structure.
If the eigenvalues of the Majorana mass matrix do not have hierarchy,
the neutrino mass after seesaw will be the squared-hierarchy,
for example, $m_{\nu_2} : m_{\nu_3} \simeq m_c^2 : m_t^2$.
Thus, considering recent experiment of atmospheric and solar neutrino oscillation,
we need a hierarchical structure in the Majorana mass matrix.
If the right-handed leptons are localized at $y=\pi R$ and the brane interaction is introduced 
only at $y=0$,
the Majorana mass matrix can have a hierarchical structure.
The large mixing angles for neutrino oscillation
can be originated from this Majorana mass structure.
We note that the 4-th generation lepton mass (which we call $a$ in the quark sector)
should be large enough, so the 4-th neutrino mass after seesaw should be higher than weak scale.

Finally, we will make a comment about unification of gauge and Yukawa couplings\footnote{
The numerical calculation of gauge and Yukawa unification in a 4D model
is demonstrated in the Ref.\cite{Chkareuli:1998wi}.}.
Noting that the effective top Yukawa coupling in 4D is always smaller than
4D gauge coupling at $1/R$ scale,
we have to consider additional brane-localized gauge kinetic terms
or the RG evolution of the $SU(2)_L$ gauge coupling and top Yukawa coupling.
We mention about the latter case.
Since left-right models has one more set of Higgs doublets rather than MSSM
and we have (heavy) 4-th generation in our model building,
the absolute value of beta function for $SU(2)_L$ is larger than MSSM.
Thus, we can easily make the gauge coupling at $1/R$ large enough
to produce the observed value of the top quark mass.
We make one more comment about bottom-tau ratio.
Since down-type quark and charged lepton are not unified at this stage,
there is no reason Yukawa couplings for bottom and tau are unified 
at $1/R$ scale.
But still, it is interesting that we have a possibility 
to realize the bottom-tau ratio,
since all the 5D Yukawa couplings are unified.

\section{Possible Unified Models}

In this section, we will consider possible unified model of the $SU(4)_w$ theory
which we have discussed.
Our model fits nicely in a $SO(12)$ unification.
It is well known that ordinary 4D left-right symmetric model
can be embedded in the $SO(10)$.
However, our left-right model cannot embedded in $SO(10)$
since the gauge symmetry of our model is $SU(3)_c \times SU(4)_w \times U(1)_{B-L}$,
which rank is 6.
Furthermore, the $SO(10)$ adjoint $\mathbf{45}$ does not include
uncolored bi-doublet Higgs,
and the $\mathbf{45}$ does not include MSSM doublet Higgs neither.
Thus, we can not construct the gauge-Higgs unification in $SO(10)$ scheme.
We also comment that $E_6$ adjoint includes the MSSM Higgs doublets,
but does not include bi-doublet for left-right models.

Since the symmetry $SU(3)_c \times U(1)_{B-L}$ can be easily unified to the $SU(4)_c$
as in the Pati-Salam Model \cite{Pati:1974yy},
we can consider gauge-Higgs unification scenario in $SU(4)_c \times SU(4)_w$ model\footnote{
The supersymmetric $SU(4)_c \times SU(4)_w$ in 5D is also discussed in the Ref.\cite{Shafi:2002ck}.}.
The bi-doublet Higgs fields are included in the 4D $N=2$ gauge multiplet of $SU(4)_w$ again,
and the boundary conditions are given in the same way as in Eq.(\ref{boundary-condition}).
All quarks and leptons are embedded in the hypermultiplet of $(\mathbf{4,4})$ representation.
To cancel the $U(1)_X$ anomaly, we need both $(\mathbf{4,4})$ and ($\mathbf{4,4^*}$) 
in the same way as we have discussed before, and the number of generation of the bulk fields
should be again even.
Those bulk matter multiplets do not include any unwanted fields.
After orbifolding, the theory become Pati-Salam model (with $U(1)_X$ symmetry),
and the quarks and leptons are unified into one multiplet in each chirality.
We can also break $SU(4)_c$ down to $SU(3)_c \times U(1)_{B-L}$ through boundary condition,
if we choose $P=(1,1,1,1)$ and $P^\prime = (1,1,1,-1)$ with respect to $SU(4)_c$ gauge space.
Here we take $P \neq P^\prime$ since we do not want to make colored Higgs massless.

The $SU(4)_c \times SU(4)_w$ gauge theory can be unified into a $SO(12)$ gauge theory.
In order to realize the gauge-Higgs unification in the 5D $SO(12)$ theory,
we use the boundary conditions for 4D $N=2$ gauge multiplet $(V,\Sigma)$
\begin{equation}
\left( \begin{array}{c}
         V \\ \Sigma
       \end{array}
\right) (x^\mu, -y) =
\left( \begin{array}{c}
         P V P^{-1} \\ -P \Sigma P^{-1}
       \end{array}
\right) (x^\mu, y),
\quad
\left( \begin{array}{c}
         V \\ \Sigma
       \end{array}
\right) (x^\mu, y+2\pi R) =
\left( \begin{array}{c}
         T V T^{-1} \\ T \Sigma T^{-1}
       \end{array}
\right) (x^\mu, y),
\label{boundary-condition2}
\end{equation}
and for $N=2$ hypermultiplets $(\Psi, \Psi^c)$,
\begin{equation}
\Psi (x^\mu, -y) =  P \Psi (x^\mu, y), \quad \Psi^c (x^\mu, -y) = -  P \Psi^c (x^\mu, y),
\end{equation}
\begin{equation}
\Psi (x^\mu, y+2\pi R) =  T \Psi (x^\mu, y), \quad \Psi^c (x^\mu, y+2\pi R) = T \Psi^c (x^\mu, y).
\end{equation}
We take that $P$ breaks $SO(12)$ down to $SO(10) \times U(1)_X$, and $T$ breaks
$SO(12)$ down to $SU(4)_c \times SU(4)_w$.
The adjoint representation $\mathbf{66}$ of $SO(12)$ is decomposed as
\begin{equation}
\begin{array}{ll}
\mathbf{66} = \mathbf{1}_0 + \mathbf{45}_0 + \mathbf{10}_2 + \mathbf{10}_{-2}, & \quad SO(10)\times U(1)_X, \\
\mathbf{66} = (\mathbf{15,1}) + (\mathbf{1,15}) + (\mathbf{6,6}), & \quad SU(4)_c \times SU(4)_w.
\end{array}
\end{equation}
For the $N=1$ vector multiplet $V$, the $P$ and $T$ should be assigned as 
$P,T=+$ for unbroken generators and $P,T=-$ for the broken ones. 
Since the gauge fields for the broken generator become massive for each decomposition,
the $SO(12)$ is broken down to $SU(4)_c \times SU(2)_L \times SU(2)_R \times U(1)_X$ as a result.
The vector multiplet $V$ and chiral scalar field $\Sigma$ are decomposed
into the Pati-Salam gauge symmetry (with $U(1)_X$) with $(P,T)$ signature in the following:
\begin{eqnarray}
V &=& \mathbf{(1,1,1)}_0^{(+,+)} + \mathbf{(1,3,1)}_0^{(+,+)} + \mathbf{(1,1,3)}_0^{(+,+)} 
+ \mathbf{(6,2,2)}_0^{(+,-)} + \mathbf{(15,1,1)}_0^{(+,+)} \nonumber \\
&& + \mathbf{(1,2,2)}_2^{(-,+)} + \mathbf{(1,2,2)}_{-2}^{(-,+)} 
+ \mathbf{(6,1,1)}_2^{(-,-)} + \mathbf{(6,1,1)}_{-2}^{(-,-)}, \\
\Sigma &=& \mathbf{(1,1,1)}_0^{(-,+)} + \mathbf{(1,3,1)}_0^{(-,+)} + \mathbf{(1,1,3)}_0^{(-,+)} 
+ \mathbf{(6,2,2)}_0^{(-,-)} + \mathbf{(15,1,1)}_0^{(-,+)} \nonumber \\
&& + \mathbf{(1,2,2)}_2^{(+,+)} + \mathbf{(1,2,2)}_{-2}^{(+,+)} 
+ \mathbf{(6,1,1)}_2^{(+,-)} + \mathbf{(6,1,1)}_{-2}^{(+,-)}.
\end{eqnarray}
Then we find that, for the chiral superfield $\Sigma$, only $SU(4)_c$-singlet bi-doublets have 
zero modes and those bi-doublets can be identified as low energy Higgs fields and their Yukawa
couplings with matter fields come from gauge interaction in the same way
as section 2.
The matter hypermultiplets are given as $\mathbf{32}$ and $\mathbf{32^\prime}$ representation
of $SO(12)$.
Those representations are decomposed into
\begin{equation}
\begin{array}{ll}
\mathbf{32} = \mathbf{16}_1 + \mathbf{16^*}_{-1}, & \quad SO(10)\times U(1)_X, \\
\mathbf{32} = \mathbf{(4,4)} + \mathbf{(4^*,4^*)}, & \quad SU(4)_c\times SU(4)_w, \\
\end{array}
\end{equation}
\begin{equation}
\begin{array}{ll}
\mathbf{32^\prime} = \mathbf{16}_{-1} + \mathbf{16^*}_{1}, & \quad SO(10)\times U(1)_X, \\
\mathbf{32^\prime} = \mathbf{(4,4^*)} + \mathbf{(4,4^*)}, & \quad SU(4)_c\times SU(4)_w. \\
\end{array}
\end{equation}
Taking the $P$ and $T$ appropriately, the hypermultiplets $(\Psi, \Psi^c)$ for $\mathbf{32}$ are decomposed
into $SU(4)_c \times SU(2)_L \times SU(2)_R \times U(1)_X$ as
\begin{eqnarray}
\Psi_{\mathbf{32}} &=& \mathbf{(4,2,1)}_1^{(+,+)} + \mathbf{(4^*,1,2)}_1^{(+,-)} 
+ \mathbf{(4^*,2,1)}_{-1}^{(-,-)} + \mathbf{(4,1,2)}_{-1}^{(-,+)}, \\
\Psi^c_{\mathbf{32}} &=& \mathbf{(4^*,2,1)}_{-1}^{(-,+)} + \mathbf{(4,1,2)}_{-1}^{(-,-)} 
+ \mathbf{(4,2,1)}_{1}^{(+,-)} + \mathbf{(4^*,1,2)}_{1}^{(+,+)},
\end{eqnarray}
and for the hypermultiplet $\mathbf{32^\prime}$
\begin{eqnarray}
\Psi_{\mathbf{32^\prime}} &=& \mathbf{(4,2,1)}_{-1}^{(+,+)} + \mathbf{(4^*,1,2)}_{-1}^{(+,-)} 
+ \mathbf{(4^*,2,1)}_{1}^{(-,-)} + \mathbf{(4,1,2)}_{1}^{(-,+)}, \\
\Psi^c_{\mathbf{32^\prime}} &=& \mathbf{(4^*,2,1)}_{1}^{(-,+)} + \mathbf{(4,1,2)}_{1}^{(-,-)} 
+ \mathbf{(4,2,1)}_{-1}^{(+,-)} + \mathbf{(4^*,1,2)}_{-1}^{(+,+)}.
\end{eqnarray}
Then we find that only $\mathbf{(4,2,1)}_{\pm 1}$ and $\mathbf{(4^*,1,2)}_{\pm 1}$ 
in which all the left and right-handed quarks and leptons are included have massless zero modes.
Here again, we encounter the $U(1)_X$ anomaly if we have only $\mathbf{32}$ but not $\mathbf{32^\prime}$,
vice versa.
Since each $\mathbf{32}$ and $\mathbf{32^\prime}$ gives one generation of fermions,
the number of generation which comes from bulk hypermultiplets is even number as a result.
The generation mixings for quark and lepton are introduced by adding brane fields 
in the same way as section 3.
We note that this $SO(12)$ unification gives one version of realization of gauge-Higgs unification
in the 4D $SO(10)$ gauge theory.
The $SO(10) \times U(1)_X$ decomposition of $SO(12)$ adjoint $\mathbf{66}$ includes
two vector representation $\mathbf{10}$, which can be identified to
the Higgs fields in the 4D $SO(10)$ unified models.
The $\mathbf{10}$'s include bi-doublets Higgses for left-right model 
(or we should say Pati-Salam model),
and the colored partners are projected out by using another boundary condition $T$.
Since extra $U(1)_X$ symmetry does not mix with electroweak gauge sector,
the weak mixing angle prediction is same as in ordinary $SO(10)$ theory that the gauge symmetry
is broken down to Pati-Salam symmetry,
if we neglect the brane-localized gauge kinetic terms with
large cutoff scale.
Though all the fermions are unified in spinor representations, 
the left and right-handed fermions are separated to the non-conjugated and conjugated 
chiral superfields $\Psi$ and $\Psi^c$ in our construction.

We make a comment about another possible unified gauge group 
of our $SU(3)_c \times SU(4)_w \times U(1)_{B-L}$.
One can consider the unified gauge group such as $SU(7)$ or $SU(8)$.
However, those extensions of our model are not straightforward,
since we have to care about $U(1)_X$ anomaly.
We have to have both $\mathbf{(3,4)}_{1/3}$ and $\mathbf{(3,4^*)}_{1/3}$
for the quark representation to cancel the 4D $U(1)_X$ anomaly,
but it seems difficult to make both
with correct $B-L$ charges in the decomposition of $SU(7)$ representation.
In $SU(8)$ case, we don't have to care about $B-L$ charge,
but still it is difficult to make anomaly free set of zero modes.
For all that, we can make the $U(1)_X$ anomaly free by introducing appropriate
brane fields,
but that is not a simple extension of our $SU(4)_w$ models.

\section{Conclusion and Discussion}

We have considered the 5D supersymmetric $SU(4)_w$ gauge theories.
The gauge group $SU(4)_w$ is broken down to left-right symmetric gauge 
symmetry $SU(2)_L \times SU(2)_R \times U(1)_X$ by orbifold boundary conditions,
and the model is reduced to supersymmetric left-right model (or Pati-Salam model) 
with extra $U(1)_X$ in 4D.
In building the model,
we employ a scenario of gauge-Higgs unification suggested in Ref.\cite{Burdman:2002se}.
In this scenario, the Higgs fields which break electroweak gauge symmetry
are unified to the gauge sector in the 4D $N=2$ supermultiplet.
Furthermore,
the 4D Yukawa interactions, which we need in order to give masses
to the fermions by Higgs mechanism, arise from gauge interaction in 5D lagrangian.
This is the most interesting feature of this type of gauge-Higgs unification.
The smallness of the 4D Yukawa coupling constants for 1st and 2nd generations
can be understood by the fermion localization along the 5D coordinate:
the left and right-chiral fermions are separated to the different 4D walls.
Since the localization of the fermion gives us a exponential profile
with respect to the 5D coordinate,
it is easy to realize the smallness of the coupling for the 1st family.
The model-building, in which we employ that the Yukawa coupling originates from
gauge coupling in 5D, leads us naturally to the world in which
the standard gauge group in 4D is unified in larger gauge group in higher dimension.

The usual left-right symmetry, which is embedded in $SO(10)$ gauge symmetry for example,
helps us to understand why CKM matrix is close to the identity matrix.
Actually, the CKM matrix is exactly same as identity matrix
with only one Higgs bi-doublet in supersymmetric left-right model
(only one $\mathbf{10}$ Higgs in $SO(10)$ model).
Furthermore, the quark mass ratios are same between up and down-type quark,
for example, $m_c/m_t = m_s/m_b$, 
if we have only one Higgs multiplet in the models.
Since those things do not match to observation,
we have to add at least one more Higgs into the models.
It is interesting that 
two Higgs multiplets are contained in the gauge-Higgs unification scenario.
However, in ordinary 4D construction of the left-right models with
two Higgs multiplets,
we spoil the reason that the CKM matrix is close to identity matrix
if the two Yukawa couplings with two different Higgs are general.
Thus, many people consider texture assumptions of the Yukawa couplings
or flavor symmetries in order to justify the small CKM angles.
In our model in this paper,
we don't need such a texture or flavor symmetry
to explain the smallness of the CKM angles as we have seen in section 3.
The reason is in the following:
The 4D Yukawa interaction which arise from gauge interaction in 5D
is flavor diagonal.
Though the generation mixing is caused from brane interaction,
the brane interaction does not make large mixing angle to the CKM
matrix because of the left-right symmetry
even if the generation mixing of the bulk hypermultiplets is large
for each up and down-type quarks.
Eventually, the CKM matrix is close to identity matrix,
as long as the quark masses are hierarchical.
The quark mass hierarchy can be obtained from fermion localization.
This is one of the interesting feature of our left-right model
with gauge-Higgs unification.

One more interesting feature of our $SU(4)_w$ model is
that particle contents of the zero modes in the matter hypermultiplets are
enough and sufficient
to construct 4D left-right symmetric model (or Pati-Salam model).
We don't have to add brane fields to cancel 4D gauge anomaly.
Instead, each fermion family has to have a pair with opposite $U(1)_X$ charge
to cancel 4D $U(1)_X$ gauge anomaly.
This leads to the number of families of bulk hypermultiplet to be even number.
In order to make family mixing,
we have to have at least one family of brane fields for quarks and leptons.
The brane fields fold Dirac masses with bulk hypermultiplets
and the number of families at low energy is equal to the
difference between the number of bulk hypermultiplets and the number of brane superfields.
Thus, in the minimal choice of particle contents, 
the model {\it with family mixings} has 3 families in the weak scale
(We obtain the 3-family as the minimal model with family mixing).

Finally, we comment about $U(1)_R$ symmetry.
The 5D $N=1$ supersymmetry corresponds to 4D $N=2$ supersymmetry,
and the boundary condition breaks the $N=2$ supersymmetry down to
$N=1$ supersymmetry.
The $U(1)_R$ symmetry can arise as a subgroup of
the $SU(2)_R$ in $N=2$ supersymmetry algebra.
We can assign the $U(1)_R$ charge to the 4D fields appropriately.
The $U(1)_R$ symmetry can prohibit 
the dimension five proton decay operators, and of course,
there is no dimension four operator in left-right model.


\section*{Acknowledgments}

We thank K.S. Babu, N. Haba, J. Lykken and S. Raby
for useful discussions.
Y.M. acknowledges the warm hospitality and support
of the KEK Theory Group during his visit there. 
This work was supported in part by US DOE Grants \# DE-FG030-98ER-41076
and DE-FG-02-01ER-45684.


\begin{thebibliography}{99}
%
%
%
%
%
%
%


\bibitem{Manton:1979kb}
N.~S.~Manton,
Nucl.\ Phys.\ B {\bf 158}, 141 (1979);

D.~B.~Fairlie,
J.\ Phys.\ G {\bf 5}, L55 (1979);
%
Phys.\ Lett.\ B {\bf 82}, 97 (1979);

G.~Chapline and R.~Slansky,
Nucl.\ Phys.\ B {\bf 209}, 461 (1982).


%
%

\bibitem{Hosotani:1983xw}
Y.~Hosotani,
Phys.\ Lett.\ B {\bf 126}, 309 (1983);
%
Phys.\ Lett.\ B {\bf 129}, 193 (1983);
%
Phys.\ Rev.\ D {\bf 29}, 731 (1984);
%
Annals Phys.\  {\bf 190}, 233 (1989).

%
%
%
%
%
%

\bibitem{Hall:2001tn}
L.~J.~Hall, H.~Murayama and Y.~Nomura,
Nucl.\ Phys.\ B {\bf 645}, 85 (2002)
[arXiv:hep-th/0107245];

R.~Dermisek, S.~Raby and S.~Nandi,
Nucl.\ Phys.\ B {\bf 641}, 327 (2002)
[arXiv:hep-th/0205122].


%
%

\bibitem{Krasnikov:dt}
I.~Antoniadis,
Phys.\ Lett.\ B {\bf 246}, 377 (1990);

N.~V.~Krasnikov,
Phys.\ Lett.\ B {\bf 273}, 246 (1991);

I.~Antoniadis and K.~Benakli,
Phys.\ Lett.\ B {\bf 326}, 69 (1994)
[arXiv:hep-th/9310151];

H.~Hatanaka, T.~Inami and C.~S.~Lim,
Mod.\ Phys.\ Lett.\ A {\bf 13}, 2601 (1998)
[arXiv:hep-th/9805067];

G.~R.~Dvali, S.~Randjbar-Daemi and R.~Tabbash,
Phys.\ Rev.\ D {\bf 65}, 064021 (2002)
[arXiv:hep-ph/0102307];

N.~Arkani-Hamed, A.~G.~Cohen and H.~Georgi,
Phys.\ Lett.\ B {\bf 513}, 232 (2001)
[arXiv:hep-ph/0105239];

I.~Antoniadis, K.~Benakli and M.~Quiros,
New J.\ Phys.\  {\bf 3}, 20 (2001)
[arXiv:hep-th/0108005];

C.~Csaki, C.~Grojean and H.~Murayama,
arXiv:hep-ph/0210133.


%
%

\bibitem{Hall:2001zb}
L.~J.~Hall, Y.~Nomura and D.~R.~Smith,
Nucl.\ Phys.\ B {\bf 639}, 307 (2002)
[arXiv:hep-ph/0107331].

%
%

\bibitem{Burdman:2002se}
G.~Burdman and Y.~Nomura,
arXiv:hep-ph/0210257.


%
%
%
%
%


\bibitem{Haba:2002vc}
N.~Haba and Y.~Shimizu,
arXiv:hep-ph/0212166.


%
%


\bibitem{Scherk:1978ta}
J.~Scherk and J.~H.~Schwarz,
Phys.\ Lett.\ B {\bf 82}, 60 (1979);
%
Nucl.\ Phys.\ B {\bf 153}, 61 (1979);

E.~Witten,
Nucl.\ Phys.\ B {\bf 258}, 75 (1985);

P.~Candelas, G.~T.~Horowitz, A.~Strominger and E.~Witten,
Nucl.\ Phys.\ B {\bf 258}, 46 (1985);

%
L.~Dixon, J.~Harvey, C.~Vafa and E.~Witten,
%
Nucl.\ Phys.\ B {\bf 261}, 651 (1985).



%
%
%


\bibitem{Kawamura:1999nj}
Y.~Kawamura,
Prog.\ Theor.\ Phys.\  {\bf 103}, 613 (2000)
[arXiv:hep-ph/9902423];
%
Prog.\ Theor.\ Phys.\  {\bf 105}, 999 (2001)
[arXiv:hep-ph/0012125];
%
Prog.\ Theor.\ Phys.\  {\bf 105}, 691 (2001)
[arXiv:hep-ph/0012352].







%
\bibitem{Hebecker:2001jb}
{\it see for example}

G.~Altarelli and F.~Feruglio,
%
Phys.\ Lett.\  B {\bf 511}, 257 (2001)
[arXiv:hep-ph/0102301];
%

A.~B.~Kobakhidze,
Phys.\ Lett.\ B {\bf 514}, 131 (2001)
[arXiv:hep-ph/0102323];
%

L.~J.~Hall and Y.~Nomura,
Phys.\ Rev.\ D {\bf 64}, 055003 (2001)
[arXiv:hep-ph/0103125];
Phys.\ Rev.\ D {\bf 66}, 075004 (2002)
[arXiv:hep-ph/0205067];

A.~Hebecker and J.~March-Russel, 
%
Nucl.\ Phys.\ B{\bf 613}, 3 (2001)
[arXiv:hep-ph/0106166];
%
Nucl.\ Phys.\ B {\bf 625}, 128 (2002)
[arXiv:hep-ph/0107039];

C.~Csaki, G.~D.~Kribs and J.~Terning,
Phys.\ Rev.\ D {\bf 65}, 015004 (2002)
[arXiv:hep-ph/0107266];

N.~Maru,
Phys.\ Lett.\ B {\bf 522}, 117 (2001)
[arXiv:hep-ph/0108002];

L.~J.~Hall, Y.~Nomura, T.~Okui and D.~R.~Smith,
Phys.\ Rev.\ D {\bf 65}, 035008 (2002)
[arXiv:hep-ph/0108071];

R.~Dermisek and A.~Mafi,
Phys.\ Rev.\ D {\bf 65}, 055002 (2002)
[arXiv:hep-ph/0108139];

T.~Watari and T.~Yanagida,
Phys.\ Lett.\ B {\bf 519}, 164 (2001)
[arXiv:hep-ph/0108152];

K.~S.~Babu, S.~M.~Barr and B.~s.~Kyae,
Phys.\ Rev.\ D {\bf 65}, 115008 (2002)
[arXiv:hep-ph/0202178];

F.~Paccetti Correia, M.~G.~Schmidt and Z.~Tavartkiladze,
Nucl.\ Phys.\ B {\bf 649}, 39 (2003)
[arXiv:hep-ph/0204080];
%
Phys.\ Lett.\ B {\bf 545}, 153 (2002)
[arXiv:hep-ph/0206307];

N.~Haba and Y.~Shimizu,
arXiv:hep-ph/0210146;

%
%
%


Y.~Mimura and S.~Nandi,
Phys.\ Lett.\ B {\bf 538}, 406 (2002)
[arXiv:hep-ph/0203126];

R.~N.~Mohapatra and A.~Perez-Lorenzana,
Phys.\ Rev.\ D {\bf 66}, 035005 (2002)
[arXiv:hep-ph/0205347];

I.~Gogoladze, Y.~Mimura and S.~Nandi,
arXiv:hep-ph/0210320,
to be published in Phys.Lett.B.

%
%
%
%
%
%


\bibitem{Pati:1974yy}
J.~C.~Pati and A.~Salam,
Phys.\ Rev.\ D {\bf 10}, 275 (1974).


%
%
%



\bibitem{Mohapatra:1974hk}
R.~N.~Mohapatra and J.~C.~Pati,
Phys.\ Rev.\ D {\bf 11}, 566, 2558 (1975);

G.~Senjanovic and R.~N.~Mohapatra,
Phys.\ Rev.\ D {\bf 12}, 1502 (1975);

%
%

R.~N.~Mohapatra and G.~Senjanovic,
Phys.\ Rev.\ Lett.\  {\bf 44}, 912 (1980);
%
Phys.\ Rev.\ D {\bf 23}, 165 (1981);

G.~Senjanovic,
Nucl.\ Phys.\ B {\bf 153}, 334 (1979).







%
%
%
%
%
%

\bibitem{Arkani-Hamed:2001tb}
N.~Arkani-Hamed, T.~Gregoire and J.~Wacker,
JHEP {\bf 0203}, 055 (2002)
[arXiv:hep-th/0101233].


%
%
%
%
%
%


\bibitem{Chkareuli:1998wi}
J.~L.~Chkareuli and I.~G.~Gogoladze,
Phys.\ Rev.\ D {\bf 58}, 055011 (1998)
[arXiv:hep-ph/9803335].


%
%
%
%

\bibitem{Shafi:2002ck}
Q.~Shafi and Z.~Tavartkiladze,
Phys.\ Rev.\ D {\bf 66}, 115002 (2002).





\end{thebibliography}
\end{document}